# Finding and Recommending Scholarly Articles

Michael J. Kurtz and Edwin A. Henneken
Harvard-Smithsonian Center for Astrophysics

## 0. Introduction

Communication is a two way street; someone speaks, and someone else listens. Scholarly communication is essentially the same, but given the growing cacophony of very many voices, who should one listen to?

The rate at which scholarly literature is being produced, similar to the rate at which all things of economic value (as measured by world GDP) are being produced, has been increasing at approximately 3.5% per year for decades (estimated using data from the World Bank and the Web of Science). This means that during a typical 40 year career the amount of new literature produced each year increases by a factor of four. The overall growth of available digital information is even more dramatic: in his IDC White Paper of 2008 (page 3), John Gantz observed that "the amount of digital information produced in the year should equal nearly 1,800 exabytes, or 10 times that produced in 2006". Over this short period of time people are not learning to read four times faster, nor are they becoming four times smarter. The methods scholars use to discover relevant literature must change. Just like everybody else involved in information discovery, scholars are confronted with information overload. Information overload is known to lead to decreased productivity and therefore has a negative economic impact, to organizations and society as a whole. Spira (2010) calculated that in 2010 information overload costs the U.S. economy a minimum of $997 billion. The opposite is true as well, of course. Kurtz et al. (2005) estimated that the ADS, by making the process of finding information more efficient, saved the astronomical community about $250 Million.

Two decades ago, this discovery process essentially consisted of paging through abstract books, talking to colleagues and librarians, and browsing journals. A time-consuming process, which could even be longer if material had to be shipped from elsewhere. Now much of this discovery process is mediated by online scholarly information systems, such as WoS, Scopus, SciFinder, NASA ADS, inSPIRE, ACM-DL, PubMed, MathSciNet, Microsoft Academic Search, Google Scholar, SSRN, RePEc, INSPEC, and numerous others, large and small.

All these systems are relatively new, and all are still changing. They all share a common goal: to provide their users with access to the literature relevant to their specific needs. To achieve this each system responds to actions by the user by displaying articles which the system judges relevant to the user's current needs. In plain English the system recommends articles for the user's consideration.

Recently search systems which use particularly sophisticated methodologies to recommend a few specific papers to the user have been called "recommender systems". Examples of successful recommender systems are Eigenfactor Recommends (Bergstrom 2012) which analyzes the citation network, and the Bx system (Bollen and Van de Sompel 2006) which analyzes the co-readership graph. Google Scholar has implemented a recommender based on the citation and authorship graphs.

These methods are in line with the current use of the term "recommender system" in computer science (e.g. Lu 2012, and Jannach 2010); typically these methods are used to recommend products, books, movies, people (for dates), music, …. In the domain of scholarly articles this sort of system is primarily intended to provide a browse capability.

We do not adopt this definition, rather we view systems like these as components in a larger whole, which is presented by the scholarly information systems themselves. In what follows we view the recommender system as an aspect of the entire information system; one which combines the massive memory capacities of the machine with the cognitive abilities of the human user to achieve a human-machine synergy.

1. Recommendation and search

From the point of view of a user of an information service search, recommendation and browse are different functions. Search might be defined as returning answers to user supplied specific questions, the equivalent to a person retrieving a fact from memory; while recommendation could be defined as returning specific answers to unstated or general questions, the equivalent to having a fact simply pop into one's head or by asking a question to an export. In this sense, a recommender system really is a technological proxy for a social process. Browsing is normally thought of, by users, as a serendipitous process, like finding a good book on a used book table, or a relevant paper in a table of contents.

From the point of view of a scholarly information service these functions are fundamentally the same. The goal of search systems, browse systems and recommender systems is to provide the user with articles relevant to his/her current needs. Optimal results are very much user dependent; it is clear that lists of articles which would best serve a beginning student would not be the same as the lists which would best serve an established expert. The increasing sophistication of search systems has fully blurred the semantic difference between search and recommendation, witness the popular aphorism: Recommendation is the new search (remark made by Stephen Green, during a talk about Minion at Harvard in 2008).

Modern information systems present the user with an information environment specific to each user. It is the task of the scholarly information system to orchestrate a continual transmutation of the scholar user's intellectual environment to both anticipate and respond to that user's unique and ever changing wants and needs. As extenders and enhancers of human thought, it is obvious that the abilities and capabilities of these machines will play a crucial role in the future development of humans.

There is a vast literature on search and recommendation systems (Lu, et al 2012); they figure prominently in the list of the world's most valuable industrial corporations. For a good collection of reviews see Ricci, et al 2010; the yearly conference on recommendation systems of the Association for Computational Machinery (recsys.acm.org) provides a snapshot of the current state of the art.

Here we will make no attempt at a comprehensive literature review, rather we will look at the general issues and techniques for recommending scholarly articles, taking examples as needed primarily from our own work with the Smithsonian/NASA Astrophysics Data System (hereafter ADS).

2. Unique problem of scholarly recommendation

Recommendations form an important part of everyday life, providing recommendations for what restaurant to visit, or which book to read, or which play to watch, which car to buy, or,..., or … They are an important feature of most newspapers.  Books and magazines, such as the Guide Michelin or Consumer Reports have a long history.  Blurring the distinction between recommending and advertising is a common and profitable practice, as the success of Google and Amazon shows.

Scholarly articles are a substantially denser and more subtly varied corpus than most commercial or entertainment domains.  For example the ADS has five times as many articles about a single subject, cosmology (a subfield of astrophysics, itself a subfield of physics, a substantially smaller discipline than chemistry or medicine) as there are films listed in the Videohound's Golden Movie Retriever or available through Netflix.

The usage pattern of scholarly articles also provides challenges.  The maximum readership rate for an article is on the day it is published or posted on-line; before any use information is available.  Use declines rapidly; the median download rate for a ten year old article from the Astrophysical Journal, the largest and most prestigious journal of astronomy, is once per month.

Citations provide another measure which can be used in developing recommendations (see for example, Küçüktunç et al. 2012), but they build up slowly.  The most cited article from the Astrophysical Journal one year past publication has typically 50-60 citations; the median article has 5.

The users of scholarly publications tend to be highly discerning individuals, typically with doctorates in the subject matters being read.  This increases the opportunities for interaction between the user and the search/recommendation system, as the user often has a detailed knowledge of what exactly s/he is seeking.  This also makes the task of pure recommendation harder, as many of the apparently "best" articles to recommend are ones the user is fully aware of, and has likely already read.

3.  The scholarly information garden

Bates (1989) has famously compared information seeking to berry picking, with the information system tasked with providing a bush filled with delicious berries.  Today's (and tomorrow's) information systems must do more than provide berries; they must create complex, interactive virtual gardens.  Besides a bush with delicious berries (e.g. a list of papers returned from a query) there are also multiple (intellectual) pathways and trail markers.

In this metaphor the search system provides the berry bush, the user interface provides the trail markers, and the recommender systems provide the pathways.  The difference between this garden and a real garden is that the topology of the information garden changes as the information system's perception of the needs of the user changes.  This continual transmutation of the garden, in response to the user's actions, makes the entire system highly interactive, a true human+machine collaboration.

Because the human is the master in the master-servant relation with the machine, there are substantial constraints on the recommender function, the portion of the system most under machine control.  When the master asks for something the servant should fetch it, not return a set of objects which the servant thinks are best. The servant could be allowed, however, to return items that other masters think are relevant (where "other masters" could be people very close in the information space, with perhaps a number of additional criteria).

As an example, a typical complex user query could be: I want to see the paper by Dressler, et al which is referenced in the recent paper on weak lensing by Kurtz, et al. This query can be asked, and the paper retrieved within about 30 seconds using the default settings of ADS, WoS, or SCOPUS. Using a recommender based system, such as Google Scholar, takes substantially longer. The most valuable scholarly commodity is the time of our top scholars; this places strong constraints on recommendation systems.

4. An Overview of Recommendation in the ADS

Recommendation techniques have been built into the ADS system since its inception in 1992; in this section we will discuss how they are implemented in the ADS, using our new, streamlined, but still in beta test, user interface system as the model. In the next section we will discuss some algorithmic details of our implementation.

In addition to the recommenders built into the search system, the ADS provides a weekly, custom (for each user) stand alone recommendation/notification service: myADS. Using a profile of the user, along with use and citation statistics, myADS provides a view of what the user should read/have read based on the papers released in the past week. There are several versions of myADS, based on input literature; the first author's version, based on data from the arXiv e-print server (Ginsparg 2011) is http://adsabs.harvard.edu/myADS/cache/267336764_PRE.html. The page contains eight short lists of articles, each a form of recommendation. As an example the upper left list on the first author's page shows the most recent papers which have cited a paper where MJK is an author. This page changes every Friday; about 50% of all working astronomers subscribe to myADS.

In developing an information retrieval system there are two basic types of recommendations: ones the user specifically requests (a query), and those the user does not specifically request. The difference between the two is not algorithmic, but lies in how they are exposed through the user interface.

The core of the ADS user interface, like many similar systems, consists of three main pages: a query page, a results/list page, and a document/abstract page. At each of these pages the system has different information concerning the current needs of the user, and the options and recommenders available on each page reflect this. Other web service designs, including for the ADS, are clearly possible; we use the three pages as a concrete example of how recommender techniques are incorporated into a scholarly information system.

**The Query Page.**

The ADS main query page contains three major elements, the query input box, the query type toggles, and the recommender pane. We describe each in turn.

Queries to the ADS system can be either free form, or entered into the query box using a highly structured query language, or a combination. Hints are provided for the most common query constructs; an auto-complete feature provides possible queries, once the user begins to fill out the box (note that this is itself a type of recommendation).

Under the query box are seven toggles, which the user chooses to direct how the system responds to the query. Choosing one of these options, and filling out the query box, initiates a user specified recommendation. The seven query types are divided into two groups, four simple sorts, and three more complex explore the field queries.

The four sort options sort the results of a query by some value. As the result of a query can be quite large (for example the query "redshift survey" returns nearly 24,000 documents) the sort functions exactly as a recommender: these on top are the ones you should look at.

Most Recent sorts the returned documents on publication date. This is the default query for the ADS, as well as for most other scholarly information systems, such as PubMed or WoS.

Most Relevant sorts on a combination of several indicators, including date, position of the query words in the document, position of the author in the author list, citation statistics and usage statistics. This type of query is popular with commercial search engines, such as Google or Bing.

Most Cited simply sorts the returned documents on their citation count, the most cited being on top.

Most Popular sorts on the number of recent downloads.

The three "Explore the Field" options differ from the simple sorts in that the documents returned do not necessarily directly answer the query; rather they use second order operators (described in detail in the next section) on truncated, sorted query lists to form lists of recommended articles.

What People are Reading returns a list of papers currently being heavily read by people in the subfield defined by the query.

What Experts are Citing shows those papers which are most heavily referenced by the most relevant papers in the field defined by the query. These papers are often not about the query at all, but concern methods used in the subfield defined by the query.

Reviews and Introductory Papers returns just that, on the subfield defined by the query. This is achieved by finding articles which cite many highly cited articles on the desired topic.

Together these seven options provide the knowledgeable user with the ability to direct the actions of the machine assistant. The resulting lists either lead the user to desired articles, or provide the basis for a more complex man-machine interaction.

The Recommender Pane shows the user lists of articles which the user did not specifically request, based on the machine's knowledge of the user and his/her recent actions. There are three different lists, the user determines which to see via a tab interface.

Perhaps the most useful list shows the day's release of papers from the arXiv e-print service (Ginsparg 2011), sorted according to the user's interest profile: the daily myADS-arXiv. This provides a (quite complete for astrophysics and other fields where arXiv is a fully integrated part) up-to-date answer to the question "what's new and of interest to me today?"

The other two lists are based on the user's recent search history. One is simply the papers the user most recently viewed, essentially a short term memory. The second shows papers which are similar (via text similarity) to papers recently viewed, and are recently released, and are very popular.

## The Results/List Page

Lists of returned items are common to nearly all search and retrieval type systems, Bing, Buy.com, Kayak, Scirus, Data.gov, ….. These pages are the means by which the user can have a detailed interaction with the system. It is from these list pages where selection criteria can be modified and refined, and where items can be chosen. In the information garden metaphor here is where the user either chooses some berries, decides to go further down a marked path, or asks the system to create a new garden.

A decade ago almost all result list pages were simply that, lists of results. Now it is common for such pages to have two or three columns, with one column devoted to the list of results. In the two column format often a column (normally on the left) provides lists of possible ways to filter/modify the list (facets), in the three column format the additional column (normally on the right) lists suggestions or recommendations which do not alter the list. The "recommendations" section of Google, Bing and Yahoo's result pages consists of advertising. In addition to allowing filtering of the publications in the results list, the facets add a layer of useful information: who are the most prolific authors in this field? who has somebody co-authored with the most? is this field still actively being publicized ? what's the ratio refereed/unrefereed?

All three designs are currently in use by digital scholarly libraries, the one panel (MathSciNet). two panel (ACM Digital Library), and three panel (PubMed). The ADS uses the two panel model; this specifically does not allow unrequested article recommendations to be shown. This design decision resulted from the master/slave view of the literature discovery process. The master (user) has requested a specific list of articles be presented, any additional lists of articles would necessarily take space away from the display of what the user has requested. For very large displays this may not be a problem, but the current trend is toward smaller displays (laptop, tablet, phone).

The list page is where the user has the most control over the recommendations returned by the system. Except for the most specific queries (e.g. get JASIST 56, 36) system responses always return documents which are ranked according to some desired criteria. These lists may be filtered, truncated, edited by the user --- then the combined properties of the documents in the (edited) list can be used to further extract information from the database.

These often complex human/machine interactions can be understood in terms of operations on multipartite graphs (Kurtz 2011) or in terms of operations on lists of attributes (Kurtz, et al 2002). These operators use the combined properties of the articles in the list to extract information from the entire data-base.

A typical example of this is: Return all papers containing the phrase "weak lensing" (done on the query page); filter these papers to only include papers concerning the cluster of galaxies Abell 383; filter these papers to only include papers which are based on data from the Hubble Space Telescope; sort these papers by date; truncate the list to include only papers less than one year old; find all the users who read one or more of these papers in the last three months; filter these users to only include persons classified by their usage patterns as probable scientists; find all the papers read by these scientist-users in the last three months; sort this list by the number of these users who read each paper and return the sorted list to the user.

By construction this list of papers shows what is currently interesting to scientists interested in HST measurements of weak lensing near the cluster of galaxies Abell 383. By making this query an interested scientist can discover what is in the collective knowledge of astronomers with similar interests. This is obviously a custom recommender which relies heavily on the synergy of man and machine. In terms of the actual implementation, the user here must type in the original phrase ("weak lensing"), click on two buttons (Abell 383 and HST), slide the date slider, and click on the most co-read button.

Most co-read is one of the recommender functions available from the list page; the What People are Reading query, on the query page, is implemented using the most co-read function. The two other "Explore the Field" options on the query page are implemented with two other recommender functions: the Get Reference Lists function is used in the What Experts are Citing option and the Get Citation Lists function is used in the Reviews and Introductory Papers option.

Get Reference Lists simply collates all the reference lists of the papers in the original list, and returns the collated list, sorted by the number of papers in the original list which referenced each paper. This gives the papers which were most cited by the articles in the original list.

Get Citation Lists collates all the lists of citing papers for each of the papers in the original list, and returns the collated list sorted by the number of papers in the original list which are cited by the citing paper. This gives the papers which have the most extensive discussions of the topic of the original list.

Another function takes all the articles in the list, combines them and treats them as a single document. It then returns the papers closest to this combination sorted by nearness according to a text similarity metric.

A final recommender function is different from the others, in that it is not intended to be used in combination with a (perhaps filtered and edited) query list. The ADS allows users to create, edit, save and recall lists. This capability is often used by authors when preparing a manuscript to prepare their bibliographies. The Citation Helper function suggests papers which are not in the list, but which are near neighbors of these papers in the citation-reference network. This is intended to help find missed references.

The facets or filters can also be viewed as suggestions, or recommendations to the user for possible additional restrictions which the user might want to make on the list. Which are the most useful possible restrictions is clearly a context specific issue. As an example typical subject matter queries can return thousands of articles with thousands of individual authors. Which ones should be on the top of the Authors facet? Simple, context sensitive metrics allow the facets for a "show me the most recent papers on weak lensing" query to be different from the facets for a "show me the most popular (or cited, or …) papers on weak lensing" query.

The Document/Abstract Page

The principal function of the Abstract page is to provide the user with a concentrated description of an article, sufficient for the user to decide whether or not to download and read it. By coming to an abstract page the user has informed the system of his/her current interest, which permits useful recommendations to be made.

In the ADS instantiation this page includes the standard metadata of title, author, journal and abstract, but it also includes links to additional information and two sets of recommenders, one user directed and one automated.

The four user specified recommendations are nearly identical to the four functions on the list page (the only difference with the result obtained from a one article list is the sort order), but are likely more familiar than their extensions.  Each of these four buttons returns a list of recommended articles.

The References button returns the list of papers referenced by the article; the Citations button returns the list of papers which cite the article; the Co-Read button returns the list of papers most read recently by probable scientists who read the article; and the Similar Articles button returns the list of articles most similar to the article by weighted text similarity.

This page also includes  a panel with eight recommended articles, clicking on any of these returns the abstract page for the recommended article.  These papers are each computed by a different algorithm (Henneken et al 2011), but each recommendation is based on a list of articles which are recently published, from major journals, and near to the original article in a vector space which we create from the index terms attached to the papers in the reference lists of all the recent major journal articles, as well as their readership pattern among professional astronomy researchers.

Taking the list of papers near to the original paper in the vector space (the near list) the eight recommended papers are:  the paper in the near list closest to the original paper; the paper which was read by the largest number of scientists who also read a paper in the near list in the last three months; the paper which was most frequently read by a scientist immediately following that person reading a paper in the near list; the paper which was most frequently read by a scientist immediately before that person read a paper in the near list; the most recent paper, among the 30 most read papers by scientists who read a paper in the near list in the last three months; the paper which the papers in the near list cite most frequently; the paper which cites the largest number of papers in the near list; and finally, the paper which refers by name to the largest number of astronomical object (stars, galaxies, …) which are also referred to by papers in the near list.  This last function uses data from the SIMBAD database (Wenger, et al 2000) of the Strasbourg (France) Data Center.

5.  Implementation techniques and details

The actual implementation of the search and discovery tools in the ADS is very simple.  We distinguish between "first order" queries and "second order" queries (Kurtz 1992, Kurtz, et al 2002).

First order queries involve search terms and a sort order, such as: "bring me the papers which contain the phrase "weak lensing" sorted by citation count."  The facets allow these queries to be modified by further filtering, such as: "restrict the list to refereed papers by Hoekstra which use either data from the Hubble Space Telescope or the European Southern Observatory and use data from the Hubble Deep Field."  The art here is not in how the queries are executed, which is pretty standard, but in making these capabilities easily available to the user.

Second order queries take lists of articles as input, and return different lists, according to attributes of the entries in the original list.  When the parameters defining the original list are well chosen, and the attributes chosen for the second order query make sense, this methodology can result in highly specific results.

Lists can either be created or edited by the user, as in the get co-read, reference, and citation list queries on the list page, or taken directly from a first order query, as are the three explore the field options on the query page. The Abell 383 example above demonstrates the degree of sophistication possible with a user intermediated query.

The Reviews and Introductory Articles option on the query page is an example of an automated chaining of a first and second order query. Again for concreteness taking "weak lensing" as the original query the procedure is: return the articles which contain the phrase "weak lensing;" sort this list by number of citations; truncate this list to get the 200 most cited articles which contain the phrase "weak lensing;" for each of these 200 articles retrieve the list of articles which cite it; collate these lists and sort the collated list by frequency of appearance so the article which cites the largest number of the 200 articles is on top..

This returns, by construction, articles which have extensive discussion sections on the desired topic, in this case "weak lensing;" normally these are review articles. The ADS has been providing this function since 1996.

Other second order operators are possible in addition to the ones we expose to the users; the procedures used in the recommender pane of the abstract page list several of them. As the multipartite network of data objects and their attributes become both richer and broader many more are likely to be developed. As an example authors are data objects which have attributes; one can imagine queries which make use of these attributes. "Show me the most popular papers among researchers who read recent papers by young authors at Japanese research institutes" is a query which some might find useful. The ADS already contains all the information necessary to answer this query, save for the age or birth year of the authors.

The ADS also uses partial match word technologies in determining the distance between documents by text similarity. Currently we are, with the kind collaboration of essentially every publisher of physics and astronomy technical literature, establishing a complete collection of the full text of nearly every article ever published in those fields. We expect these full text data will form the basis for our article similarity measures in the future, so the ones currently implemented will become obsolete.

Basically we now use two different methods: weighted word counts, modified by an extensive, subject specific synonym list (Kurtz et al 1993), a technique which has been standard since the SMART system (Salton and Lesk 1965); and an eigenvector method which uses as input the index terms associated with references in articles combined with the usage patterns of heavy users (Kurtz, et al 2010), techniques derived from Kurtz (1993), which is based on the work of Ossorio (1966).

We expect, in the near future, to build an article similarity system based on a Latent Dirichlet Allocation (Blei, Ng and Jordan 2003) of the full text of recent articles in astronomy and physics. If successful this will supplant both methods.

6. Measuring effectiveness

Scholarly communications is a substantially different domain than the normal fields of commerce where not-so-highly-interactive recommender systems are found. Amazon.com

can build a metric based on sales, Match.com on marriage rate, Netflix.com on user rankings, Google Ads on clicks, what metric could measure the effectiveness of a highly interactive scholarly information system?

Certainly there are components of an information system which can be measured by a click rate. One can easily imagine that different recommender algorithms could be compared in the recommender panel on the abstract page, for example. These suggested articles are, however, a very small component of the total service.

Ranking queries by relevance is also highly problematic. What is relevant to one person is not to another; the ADS alone has seven different types of rankings on its query page. There is no reason to believe these seven are complete or unique.

Even when a particular query type can be measured, it is not obvious what to do, as the query itself is embedded in a larger context. For example: it is reasonable to expect that a measure of "importance" can be defined, and that network based methods, such as Eigenfactor (West, et al 2009) could show significantly better performance, albeit not hugely better (Davis 2008), according to this metric than simple citation counts.

Were this a stand-alone query the discussion would end with the measurements; it is not. The query is part of a system which includes the user; the user must decide to invoke this query sorting, and may choose to use the result in a further query. It is of crucial importance that the user understands what the query is doing.

Even for the ADS' heavy users, essentially all of whom have doctorates in physics, it is much easier to understand a ranking based on citations than a ranking based on the eigenvalues of the principal eigenvector of the citation connectivity matrix (or some other similar technique).

The goal of all the scholarly informations systems is to improve the quality of research. Few would argue that these systems have not improved the research effort, but how much?

Kurtz, et al (2000) devised a method to measure the impact of the (then new) digital literature system compared with the old paper system. They assumed that in the paper era a researcher did not waste his/her time (on average) by going to the library and reading an article. Since the electronic paper is available essentially instantly, they suggested that the no longer required overhead of going to the library (photocopying the document, …) could be counted as additional research time obtained through use of the digital technology.

Kurtz, et al (2005) used this technique to find that the use of electronic documents through the ADS contributed 736 additional full-time equivalent researcher years to the world-wide astrophysics research effort, about 7% of the world's astronomers. Lesk (2011) has suggested that this could be an underestimate, as it does not take into account the improved mean quality of the papers read, due to the improved search techniques.

Scholarly information systems do not just recommend articles. As an example the Harvard Catalyst is an information system covering medical research. As one of its features it recommends possible people as collaborators. This is likely a very useful function, but, considering privacy concerns, it is difficult to imagine, even in principle, a robust methodology for determining the quality of these recommendations apart from anecdotes.

As these systems become mature there will likely be comparisons of the differences between similar techniques, with better performers replacing poorer performers. Currently these systems are still quite new and rapidly changing. Their capabilities and usefulness depends far more on the vision of their respective creators and on the quality of the offered data than on any optimization techniques.

7. Conclusions/Future

Recommendations are central to the effective use of our new digital environment. A decade ago recommendation techniques were similar across fields. The ADS, for example, offered the capacity to find papers which "cited this (these) paper(s) also cited" in 1996; somewhat later Amazon.com introduced their "people who bought this book also bought" feature. Now, however, the path of research literature has diverged from the mainstream.

In terms of information density, frequency of use, and user expectations the problem of recommendation of research literature is substantially different from the more familiar commercial applications. It is simply not possible to solve this problem by algorithmic means alone, the information is too dense, the use too infrequent, and the expectations of the researcher/user too high. Scholarly recommender systems must be close collaborations of humans with machines.

This collaboration is becoming hugely powerful. With the aid of a machine assisted memory an astrophysicist can "remember" every single word of every one of the 125,000 articles which contain the word "redshift," but which ones does s/he need to read now to facilitate his/her research? It is the ongoing task of scholarly information systems; of scholarly recommendation systems, to develop the means by which users can search and find the berries that they need in the vast memory garden now at their disposal.

Amazing progress has already been made. It is now a simple matter to find out "what is everyone else in this field reading that I should be reading?" Automatically accessing the collective knowledge of relevant subgroups of scientists smacks of science fiction (Kurtz 2011), but it is happening now routinely.

The field is still very young, major changes are certain to occur. No one over the age of 40 learned to use any of the major modern scholarly informations systems as an undergraduate, because they did not yet exist then. Nearly every principal investigator on every major grant, every full professor at every major university learned to use the research literature on paper. In the future this will no longer be true.

Information systems are becoming vastly richer, more complex, and more closely interconnected and interoperative. Articles, people, research objects, data sets, instruments, organizations, etc are now data objects, with properties and attributes which can be combined and used in a myriad of ways. It will be the ongoing responsibility of the recommender aspect of these systems to permit researchers to make the full use of these capabilities.

8. Acknowledgements

We especially thank the ADS team, led by Alberto Accomazzi. Whenever in the text we use the term "we" to refer to the ADS we actually mean the ADS team.

We acknowledge conversations with Rudi Scheiber-Kurtz, Paul Ginsparg, Herbert Van de Sompel, Carl Bergstrom, Johan Bollen, Jim Gray, Geoff Shaw, and Peter Ossorio.

The ADS is funded by NASA grant NNX12AG54G.

9. Web sites referred to in the text

adsabs.org (ADS)
webofknowledge.org (WoS)
scopus.com
pubmed.gov
eigenfactor.org
scholar.google.com
academic.research.microsoft.com
dl.acm.org
arxiv.org
cas.org/products/scifinder
ams.org/mathscinet/
www.theiet.org/inspec
ssrn.com
repec.org
inspirehep.net
simbad.u-strasbg.fr
catalyst.harvard.edu
data.worldbank.org
recsys.acm.org
viamichelin.com
netflix.com
consumerreports.org
match.com
amazon.com
google.com
movieretriever.com
bing.com
buy.com
scirus.com
kayak.com
yahoo.com
data.gov